\documentclass{acm_proc_article-sp}{
\usepackage{graphicx}
\usepackage{subcaption}
\usepackage{amsmath}
\begin{document}

\conferenceinfo{}{Bloomberg Data for Good Exchange 2017, NY, USA}

\title{Urban Explorations: Analysis of Public Park Usage using Mobile GPS Data}

\numberofauthors{1}
\author{
\alignauthor
Wenfei Xu\\
       \affaddr{CARTO}\\
       \affaddr{New York City, NY}\\
       \email{wenfei@carto.com}
}
\maketitle
\begin{abstract}
This study analyzes mobile phone data derived from 10 million daily active users across the United States to better understand the spatio-temporal activity patterns of users in Central Park, New York. The aim of this initial investigation is to create quantifiable measures for understanding public space usage in regions of the city that have no natural data source for measuring activity.  We analyze the trip behaviors of users across time and different regions in the park to find patterns of co-location and shared time and, thus, potential social interaction. We find that regions with established amenities and points of interest exhibit a higher percentage of shared experiences, indicating that institutional amenities act as 'beacons' for users' experiences in the park.

\end{abstract}

\keywords{mobile phone data, public space, mobility, hierarchical clustering}

\section{Introduction}
For numerous decades, urban designers, planners, and architects have been interested in creating a quantifiable, data-driven understanding of public space usage from both a physical design and social interaction perspective.  Data is seen as a possible aid to better and cost-effective design.  

Perhaps the two of the most well-known in this area of study are William "Holly" Whyte, who produced a \textit{The Social Life of Small Urban Spaces}\cite{whyte_social_1980}, which studies how people use and move through public spaces such as the Seagram building throughout the day, and Jan Gehl's influential \textit{Life between buildings: Using Public Space}\cite{gehl_life_2011} and the more recent Gehl Public Life Diversity Toolkit\cite{gehl_institute_public_2016} for on-the-ground data-collection. The Gehl method, which likely has the most detailed methodology, measures both physical conditions that may be more or less conducive to public space usage, and characteristics of social interactions.  Due to the time-intensive nature of physical information collection, the need to create automated, scaleable method of gathering and analyzing public space data is crucial to an improved understanding of usage.

In more recent decades, the ubiquity of mobile phones and increased availability of its data has led to the possibility that these types of surveys can be accomplished on both a larger and a more fine-grained scale.  Mobile phone data has been used to understand features such as land use, social networks in an urban context, spatio-temporal mobility, much of which contributes to our understanding of how the city is being used and where there exists potential areas of improvement. 

In the field of urban science, more broadly, mobile phone data is now widely used in studying transportation and mobility, as well as spatial social networks and land use, though it is rarely applied to the specific study of public spaces of smaller scales, such as parks.  Because public spaces often lack a natural source of data, and because they are relatively more neutral regions study for mobile phone data, they are a particularly appropriate testing ground. 

In this study, we focus on the social nature of park usage, by asking the question: What is the average percentage shared experience, and thus likelihood of interaction, with someone else? Beyond a pure count, we hope to better understand the attraction that various spatial and temporal characteristics of park usage in Central Park.

%\newpage
\section{Background}
Urban planners and designers have been long been interested in quantifiable measures of public space usage and social behavior within those spaces.  While mobile phone, and typically call-record data, has often been used to understand general patterns in human mobility \cite{gonzalez_understanding_2008} \cite{lu_approaching_2013} and patterns in the context of topics such as transportation\cite{wang_transportation_2010}\cite{alexander_origindestination_2015}\cite{becker_tale_2011}\cite{isaacman_identifying_2011} and social networks\cite{eagle_reality_2006}, research in the realm of public space usage is limited.

\subsection{Public Space Usage}

The research on public space usage has primarily been in the realm of public health. \cite{bauman_updating_2004}\cite{wendel-vos_potential_2007}. Previous research has concluded that larger public open spaces are more conducive to higher levels of walking \cite{giles-corti_increasing_2005}, while the presence of amenities such as wooded areas and different types of paths has also been shown to promote park usage.  \cite{floyd_park-based_2008}\cite{reed_descriptive_2008}\cite{kaczynski_association_2008}. More quantitative methods have been employed by researchers to understand public park usage.  For instance, using data from physical observation, McCormack et al.\cite{mccormack_physical_2014} concluded that socio-demographic environments around the park as well as physical and social characteristics within the park shape public park usage.  

The Public Life Diversity Toolkit created by the Gehl Institute, which contains various on-the-ground surveys of physical space and social interaction, represents a comprehensive method used by urban designers and planners to measure public space usage.  Nevertheless, these methods are difficult to scale.  

In recent years, there has been a wider adoption of innovative technologies to measure aspects of public space, including from companies such as Soofa and Placemeter, using sensors and computer vision, respectively. In our study, we aim to demonstrate the equal capability of data already created by cell phones to fulfill the same types of measurement needs.

\subsection{Mobile Data}

Since Eagle and Pentland's study using mobile phone data to study geographically-oriented social networks\cite{eagle_reality_2006}, the use of geo-located mobile phone data has been pervasive in the realm of urban analysis.  

In terms of the use of mobile phone data to understand land use patterns, Calabrese et al. have shown that using mobile cell phone data can be a sufficient proxy for intra-urban individual mobility at a resolution of 500m x 500m\cite{calabrese_understanding_2013}. Unsupervised clustering and 'hotspot' detection are common methods in the analysis of data: Pei et al. use a c-means clustering method to derive spatio-temporal patterns in Singapore.\cite{pei_new_2014}. Similar to this study, Louail et al. employ a non-parameterized density threshold, but set global thresholds for hotspot detection.\cite{louail_mobile_2014}.  

\subsection{Our Approach}
Our study looks specifically at Central Park using mobile phone data.  We employ a hierarchical clustering model to aggregate the data spatially and a 1-dimensional range-search algorithm to identify trips in our data.  There has been little previous quantitative analysis in the realm of urban geospatial research that looks at public space usage on the scale with which we are considering, and furthermore, little investigated in spatial multi-scale DBSCAN models. 

\section{Data}
The mobile data we work with contains traces from mobile phone applications of various kinds for iPhone and Android during the month of May, 2017.  The function of these applications include, but at not limited to, weather services, dating services, navigation, and music streaming.  For the purposes of this work, we have limited the data to those traces that are within the boundaries of New York City's public parks. 

Our study here focuses on Central Park in Manhattan, where $N$ = 556,385. Below are summary statistics for the top 5 applications in our data. 

\begin{center}
    \begin{tabular}{ | l | l | l|l|}
    \hline
    Rank & Count & Avg Acc (m.) & Acc Std.\\
    \hline
    1 & 195,404 & 46.6 & 370.8\\
    2& 109,056 &  355.8  & 648.0\\
    3& 69,020 &  202.9 & 492.1 \\
    4& 50,590 &  123.8 & 262.9\\
    5 & 47,795 & 70.2	& 149.5 \\
    \hline
    \end{tabular}
\end{center}
We remove data from the first application as it is used for routing purposes, which is evident from its traces, primarily occuring along the major routes in the park.  For the purposes of discovering clusters with a variety of activity, these data, which can have a 'gravitational' effect when we cluster, are removed from the analysis.  

Below are the main characteristics of the data with which we are concerned. For user $u$, trip $j$, cluster $k$, path $P_{j,u}$, time $t_{j,u}$, and trace $p_{i,j,u}$, we have: 
 
 \begin{itemize}

   \item \textbf{Trip Duration} For each identified trip, which we define $L_{j,u} = max(t_{j,u}) -min(t_{j,u})$
   \item \textbf{Dwell-time} For each identified trip and each identified cluster, which we define $D_{k,j,u} = max(t_{k,j,u}) -min(t_{k,j,u})$ for cluster $k$.

   \item \textbf{Velocity} For each $(p_{i,j,u},p_{i+1,j,u})$ pair of points in trip $(j,u)$, this is defined as $\frac{dist(p_{i+1,j,u},p_{i,j,u})}{(t_{i+1,j,u}-t_{i,j,u})}$.
   \item \textbf{Travel mode}.  We deduce travel from the velocity, to say that if $v_{u,j} <0.05$ then the user is in 'stay' mode. If $0.05 \leq v_{u,j} <3.1$ then the user is 'walking'. If $3.1 < v_{u,j} <10$ is 'running'.  If the trip only contains one point, we call it 'start'.
   \end{itemize}

\begin{figure}[h]
\centering
\includegraphics[width=8cm]{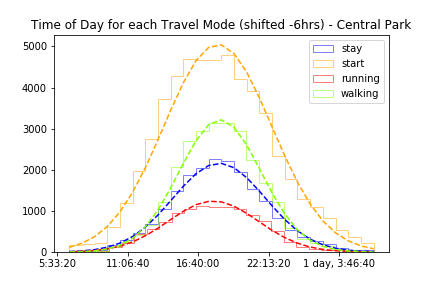}
\caption{Travel time frequency distribution - 6hr shift}
\label{travel_modes}
\end{figure}

\begin{figure}[h]
\centering
\includegraphics[width=8cm]{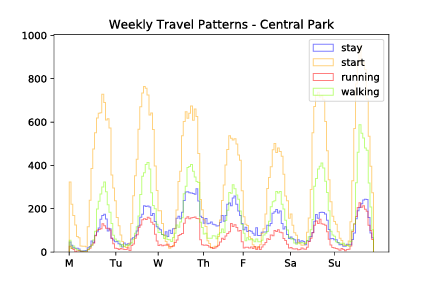}
\caption{Weekly Travel time frequency distribution}
\label{travel_modes}
\end{figure}

In figures 1 and 2, we can see that the pattern distribution of these various trip modes are fairly normally distributed throughout the day and week.  There are slightly higher volumes on the weekend versus the weekdays, with the exception of running. Despite removing the main navigation application likely used with running from our data, we still see a subset of the data that is registered as running.

\section{Method}
To help us answer questions about spatio-temporal land-use patterns in Central Park, we perform a series of analysis on the raw mobile phone data that allows us break down the data both in by individual user-trips and by spatial zones of activity.  One we are able understand these patterns, we then calculate percentage of time shared across different user-trips. 

\subsection{Temporal clustering}
Our initial step is identify individual trips based on the unicode timestamp for each user, through grouping individual points of a user that occur no more than 20 minutes from each other. So:
\begin{align*}
P_{j,u} = \{p_{i,j,u}\} \\
s.t.  \quad t_{i+1,j,u} -t_{i,j,u}< 20 min
\end{align*}
We use a wider window to allow for various applications with different ping frequencies. Figure 3 shows the distribution of trip durations for all of our data.  The median trip duration is 27.7 minutes.
\begin{figure}[h]
\centering
\includegraphics[width=8cm]{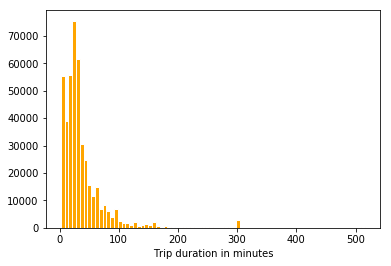}
\caption{Distribution of Trip Duration in Minutes}
\label{trip_duration}
\end{figure}

\begin{figure}[h]
\centering
\includegraphics[width=8cm]{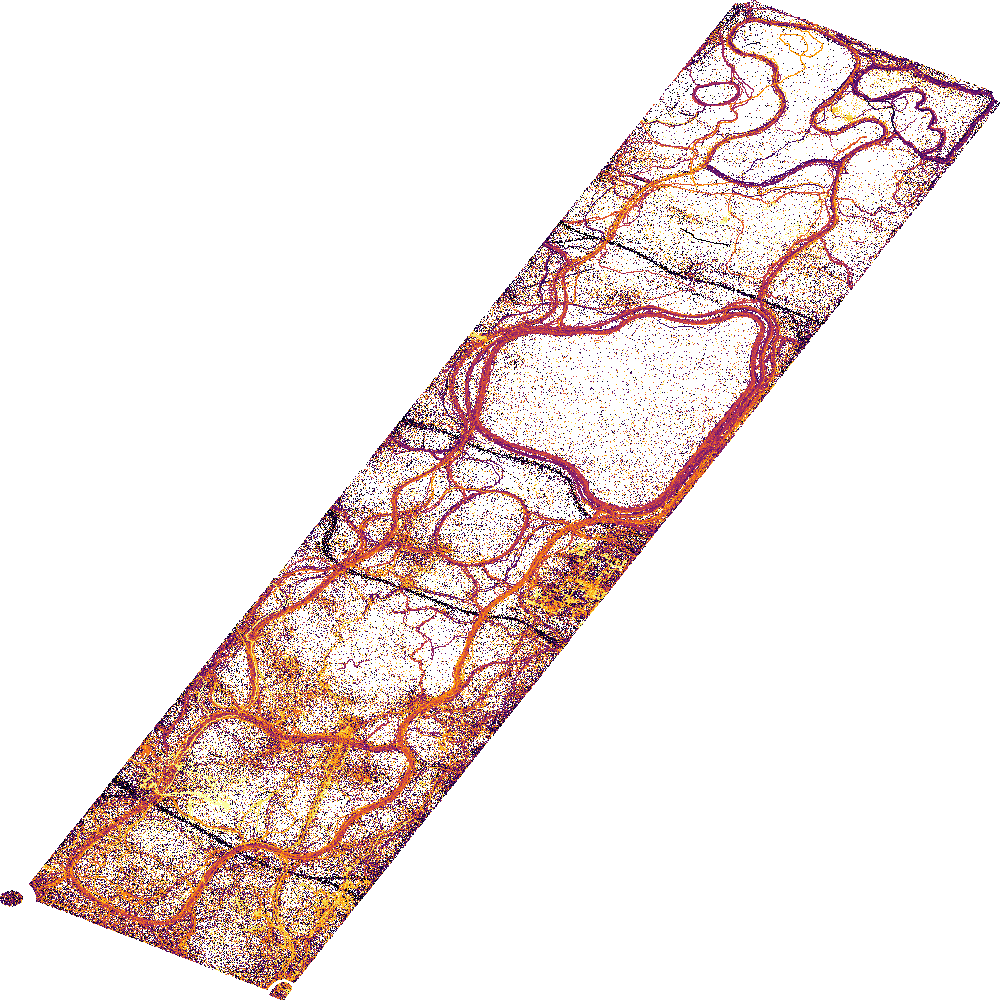}
\caption{Trip Duration (lighter color indicates longer trip)}
\label{log_velocity}
\end{figure}

\subsection{Spatial clustering}
We then cluster our data spatially in order find regions of activity within the park.  There are two motivating issues in the cluster discovery process: The first is the heterogeneous density of different types of activity in the park: for instance, a baseball game might exhibit a lower density of traces than a concert or the museum. The second is the inherently varying density of the traces from the different applications in our data.  For these reasons, algorithms such as DBSCAN, which specify a static density for all the data, would not be appropriate.  Instead we employ an approach using the hierarchical spatial clustering algorithm HDBSCAN to detect clusters\cite{campello_hierarchical_2015-1}, which removes the static density requirement and allows for various scales of clusters to occur.

\begin{figure}[h]
\centering
\includegraphics[width=8cm]{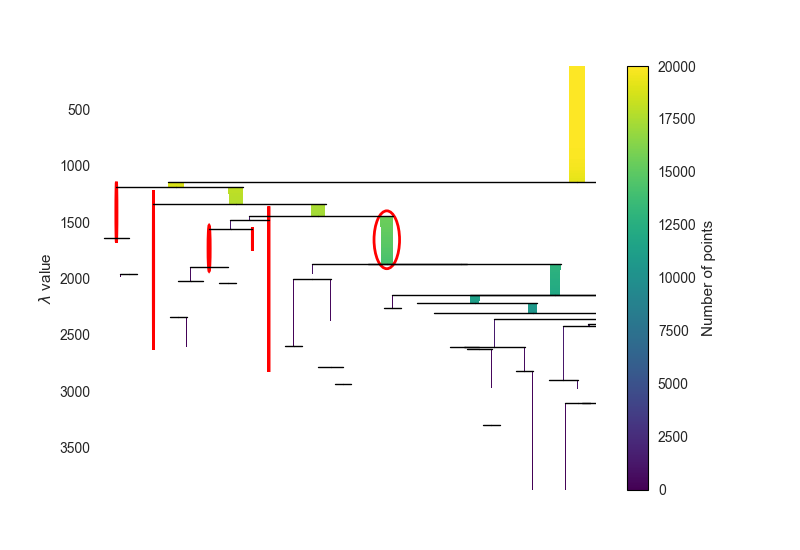}
\caption{Cluster Extraction for Hierarchical MST - Central Park}
\label{hierarchical}
\end{figure}

We use a HDBSCAN algorithm to generate clusters of varying density.  Whereas DBSCAN uses two parameters ($min_{pts}$, $\epsilon$), HBDSCAN only takes a $min_{pts}$ parameter.   We use $min_{C} = 100$.
\begin{figure}[h]
\centering
\includegraphics[width=8cm]{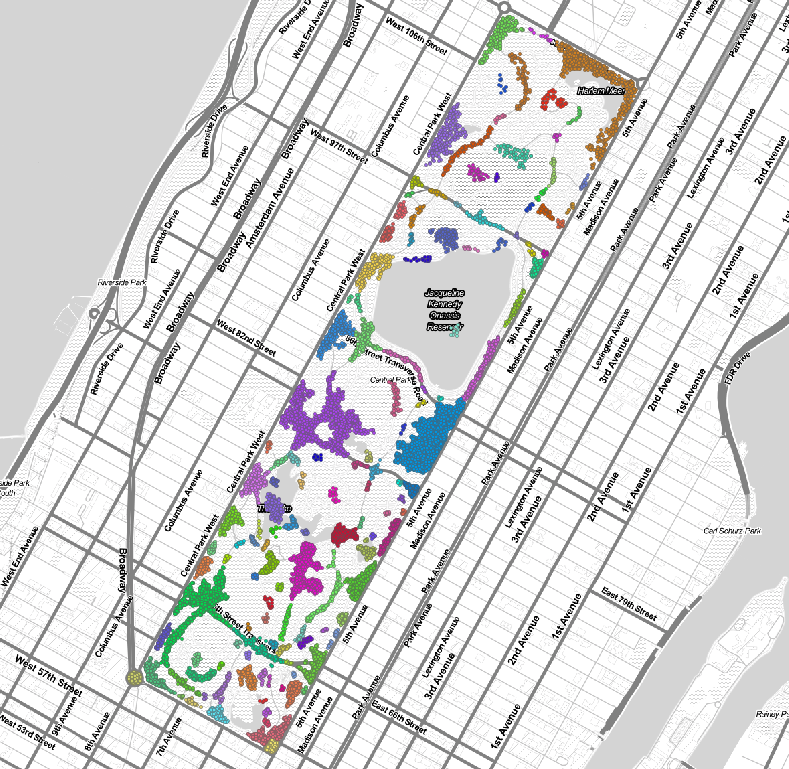}
\caption{HDBSCAN Clustering for Central Park}
\label{clustering}
\end{figure}

In fig 6 we see a sample of the resulting clustering analysis in Central Park.  We can see that it represents different regions and scales of activities.  For instance, one cluster the Huddlestone Arch in the northern region of the park, while another is the entire Metropolitan Museum of Art region. 

\begin{figure}[h]
	\centering
	\begin{subfigure}[b]{0.2\textwidth}
		\includegraphics[width=\textwidth]{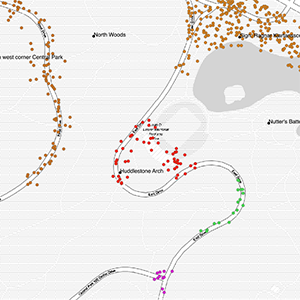}
		\caption{Huddlestone Arch}
		\label{centralpark:1}
	\end{subfigure}
	\hspace{5mm}
	\begin{subfigure}[b]{0.2\textwidth}
		\includegraphics[width=\textwidth]{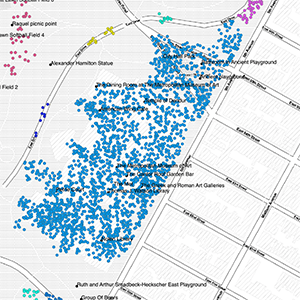}
		\caption{Metropolitan Museum}
		\label{centralpark:2}
	\end{subfigure}
\caption{Hierarchical Clustering Examples}
\end{figure}

\subsection{Shared Experience}
After creating user-trip and spatial clusters, we find the average percentage of shared user experience per cluster.  For each cluster $k$:
\begin{align*}
	P_{j1}= \frac{\sum\lim_{j}O_{j1}}{N_{J_{k}}}
	%P_{j1}= \frac{\sum\limit_{j}O_{j1}}{N_{J_{k}}}
	\text{  , for journey } j1 \text{ where}\\
	O_{j1} = D_{k,j1} \cap D_{k,j,u}\text{ and }\\
	j \in J_{k}, j \neq j1\\
\end{align*}

We'll ultimately calculate an average percentage shared experience per cluster as: 
\begin{align*}
	 P = \frac{\Sigma P_{j}}{N_{k}} 
\end{align*}
For this analysis, we remove larger clusters such as those at Columbus Circle and at the Metropolitan Museum of Art for ease of calculations.  We can make the assumption that the average percentage of shared experience of these clusters is high and follows the trend. 

\section{Results}
After we create spatial clusters and identify individual trips, we are able to better understand activity within each cluster and park usage.  In figure 9 we show the average dwell-time in minutes for each cluster against the number of unique trips on a log scale.  We can see that the distribution is roughly uniform, with possibly different regimes of distributions.   The outlier is located around the North Meadows baseball field.  

\begin{figure}[h]
\centering
\includegraphics[width=8cm]{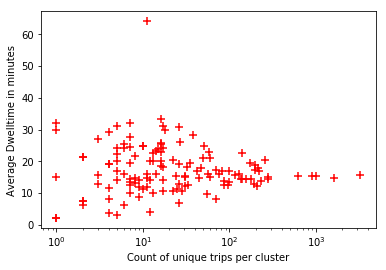}
\caption{Average Dwelltime by Cluster}
\label{dwelltime_cluster}
\end{figure}

Looking at the average percentage of shared time per user in all of the clusters, on a log-log scale, we find a linear relationship between the the number of unique trips per cluster and the number of shared interactions with a slope of approximately 0.058.  From these initial results, we can see that, while the percentage of shared time increases with larger cluster size, on average, the scaling rate is sublinear. Further investigation may show that this relationship is affected by the multi-scale nature of the various clusters in our population.  
\begin{figure}[h]
\centering
\includegraphics[width=8cm]{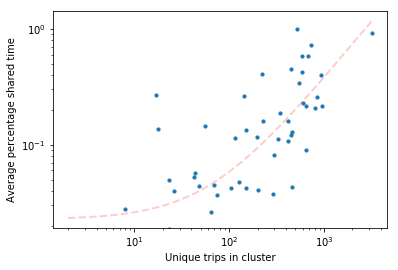}
\caption{Log-log Plot of Count and Avg. Percentage Shared Time}
\label{sharedtime}
\end{figure}

\begin{figure}[h]
\centering
\includegraphics[width=4cm]{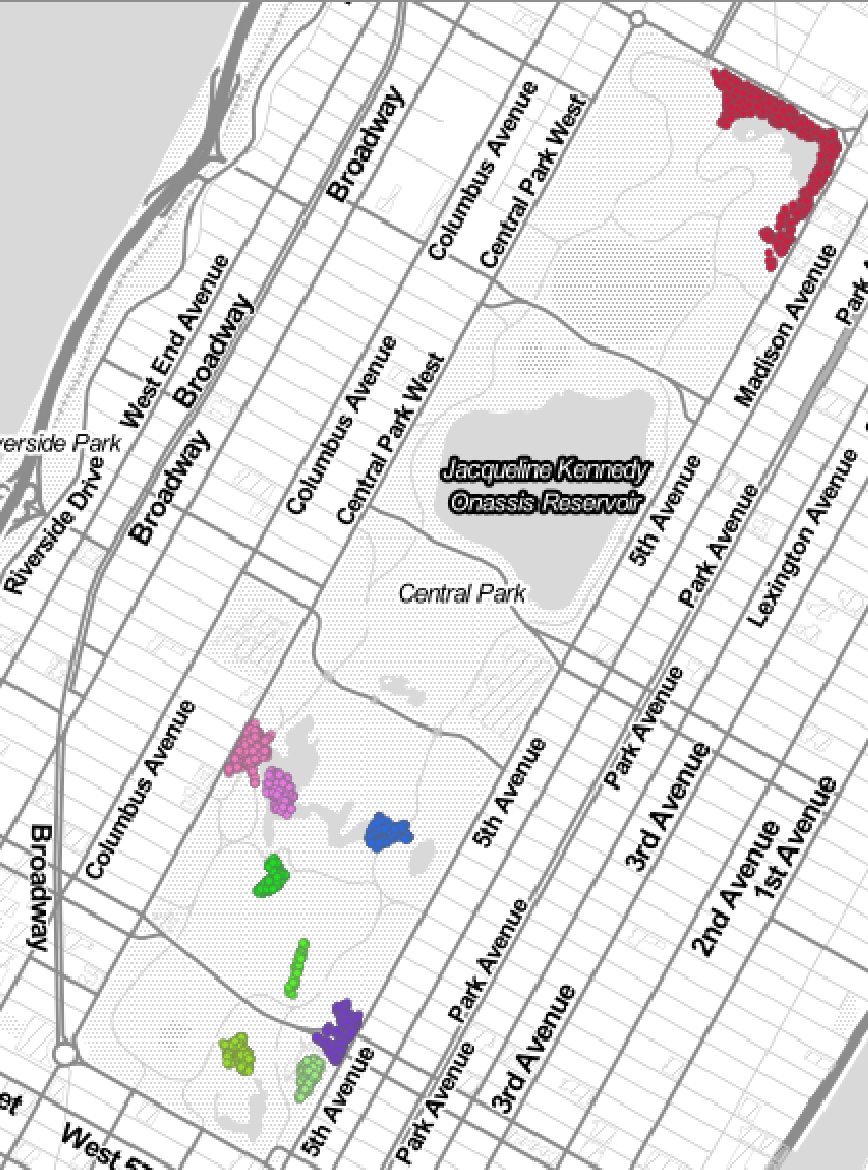}
\caption{Highest 10 Average Shared Percentage}
\label{sharedtime}
\end{figure}

Below we show the top 10 regions of average percentage of shared experience.  Some of these regions are centered around the Lake Central Park, presumably as users often share boat rides together, the Loeb Boathouse and Boathouse Bar, the Zoo, the Mall, and the Skating Rink.  

\section{Conclusion}
This research represents initial results from our research into the spatio-temporal dynamics of public space usage.  From the exercise in average percentage of shared experience, we can make some preliminary deductions about how users within the park congregate and what kinds of activities or amenities promotes more social interaction: Higher percentages of shared time appear to be occur in clusters that are associated with established amenities.  Established amenities may be a 'beacon' for users in the park. 

Further research in this topic may investigate those clusters which have much higher or lower average percentage shared experience than expected, the effect of the hierarchical clustering model the nature of each of the clusters, as well as more developed in-cluster analysis of patterns of activity.  

\section{Acknowledgments}
This research is made possible by CARTO, with data generously supplied by LiveRamp.  Special thanks to Stuart Lynn for all the help.

\nocite{*}
\bibliographystyle{abbrv}
\bibliography{references}

\end{document}